\documentclass[10pt,aps,prb,raggedbottom,longbibliography,reprint,citeautoscript,letterpaper]{revtex4-2}
\usepackage[usenames,dvipsnames]{color}
\usepackage{graphicx,microtype}
\usepackage{booktabs} 
\usepackage{multirow} 
\usepackage{amssymb}
\usepackage[bookmarks=false,colorlinks]{hyperref}
\hypersetup{linkcolor=magenta,citecolor=MidnightBlue,filecolor=Plum,urlcolor=MidnightBlue}
\usepackage[all]{hypcap} 
\usepackage[version=4]{mhchem}
\usepackage{amsmath}



\usepackage{soul}

\begin{document}

\title{Peierls-like distortion drives anion ordering in rutile TiOF} 
\author{Siddhartha S.\ Nathan}
\affiliation{Department of Materials Science and Engineering, Northwestern University, Evanston, Illinois  60208, USA}

\author{Danilo Puggioni}
\email{danilo.puggioni@northwestern.edu}
\affiliation{Department of Materials Science and Engineering, Northwestern University, Evanston, Illinois  60208, USA}

\author{James M.\ Rondinelli}
\email{jrondinelli@northwestern.edu}
\affiliation{Department of Materials Science and Engineering, Northwestern University, Evanston, Illinois  60208, USA}

\begin{abstract}
 We use first principles density functional theory calculations to examine the effect of multiple anions on the Peierls distortion in the rutile oxyfluoride TiOF. By using a structure enumeration approach, we obtain the ground state atomic structure for TiOF and identify the driving forces behind the experimentally observed two-dimensional anion  ordering along rutile (110) planes. We find that adjacent edge-connected octahedra comprise  like atoms with an –O-O/F-F/O-O/F-F– pattern along the rutile [001] direction. This anion pattern coexists with a Peierls-like distortion leading to the formation of a singlet state between neighboring Ti$^{3+}$ cations. We show that the anion ordering arises from  competition between electrostatic interactions, owing to the Ti-F cation-anion pairs, and the tendency of the $d^1$ Ti$^{3+}$ cation to form Ti-Ti dimers, characterized by increased metal-metal bonding. We find that the addition of strong on-site Coulombic interactions to the Ti $d$ manifold  suppresses the formation of the singlet state. By increasing the correlation strength, we uncover two first-order phase transitions: first, from a nonmagnetic insulator to a ferromagnetic half-metal, and then second to a ferromagnetic insulator. Last, we show that the electronic configuration of the transition metal cation in rutile oxyfluorides is responsible for the observed anion order, enabling design of ordered heteroanionic materials exhibiting collective phenomena through cation sublattice control. 
\end{abstract}

\date{\today}

\maketitle
\clearpage
\section{Introduction}

Transition metal oxides (TMO) are a topic of great interest owing to the many interesting physical phenomena they display \cite{pandey2010correlation,monthoux1991toward,moshnyaga2009electrical,PhysRevB.78.155107,puggioni2018design}.
Rational design of TMOs exhibiting collective phenomena, however, often proceeds slowly, because of the large number of competing (or at times cooperating) factors at play that produce energy minima in a shallow landscape. 
The effort is further confounded by the difficulty of disentangling intertwinned interactions. For instance, the structurally simple rutile oxide VO$_2$ undergoes a thermally driven metal-to-insulator transition (MIT) \cite{PhysRevLett.3.34}. The origin of the MIT has been a subject of prolonged debate. Different scenarios have been proposed, including a phase transition driven by structural changes due to Peierls instability \cite{woodley2008mechanism}, a pure electronic Mott-Hubbard phase transition \cite{stefanovich2000electrical}, and a hybrid of both mechanisms \cite{kim2013correlation}.In this context, heteroanionic materials (HAMs), where one or more oxygen atoms in the anion sublattice are replaced by fluorine or nitrogen, offer an opportunity to address these complexities in oxides. Moreover, harnessing multiple anions to regulate orbital occupancy, crystal field effects, and magnetic interactions presents a new avenue towards electronic property control \cite{szymanski2019design,hartman2018multiferroism}. 

Electronic structure design in heteroanionic materials (HAMs) hinges on precise control of anion ordering within heteroleptic polyhedral units and their long-range assembly \cite{charles2018structural}. Depending on the property of interest, anion disordering—whether short or long-range—can be detrimental to material performance and pose a barrier to realizing novel phenomena. Although anion order principles exist for some anion substituted materials \cite{yang2011anion,pilania2020anion}, the precise details of anion order–disorder phenomena are poorly understood both experimentally and theoretically.
In X-ray and neutron diffraction techniques,  the ability to distinguish between anions within a crystal structure depends on factors such as atomic number and coherent neutron scattering length. The larger the difference between these parameters, the easier it is to identify the occupied sites. In oxyfluorides, where O$^{2-}$ and F$^-$ have similar  atomic numbers and scattering lengths, determining anion site ordering is not straightforward \cite{krylov2014experimental,atuchin2012exploration}. This contrasts with oxysulfides, where X-Rays can differentiate O$^{2-}$ and S$^{2-}$ \cite{tsujimoto2018function} or oxynitrides, where neutron diffraction can reveal anion order \cite{yang2011anion,oka2014possible}. 
Furthermore, the spectrum between fully ordered and completely disordered anions includes numerous cases of intermediate anion orders, which may involve local clustering or extended correlations, making  the characterization of these materials challenging  \cite{zhang2011local,coles2023anion,tsujimoto2022impact}. 
To understand how local anion ordering occurs within unit cells or across various metal-anion polyhedral arrangements in oxyfluorides, electron diffraction techniques have been effective. For example, in the case of rutile oxyfluoride FeOF, anion ordering along $\langle110\rangle$ planes was attributed to the local arrangement of \textit{fac}-ordered [FeO$_3$F$_3$] octahedral units \cite{brink2000electron}. To assist and inform the synthesis of HAMs with targeted properties and their subsequent structural and physical property characterization, developing mechanistic models to understand how short and long-range anion site ordering affects the electronic structure is crucial.

\begin{figure*}
    \centering
\includegraphics[width=0.66\textwidth]{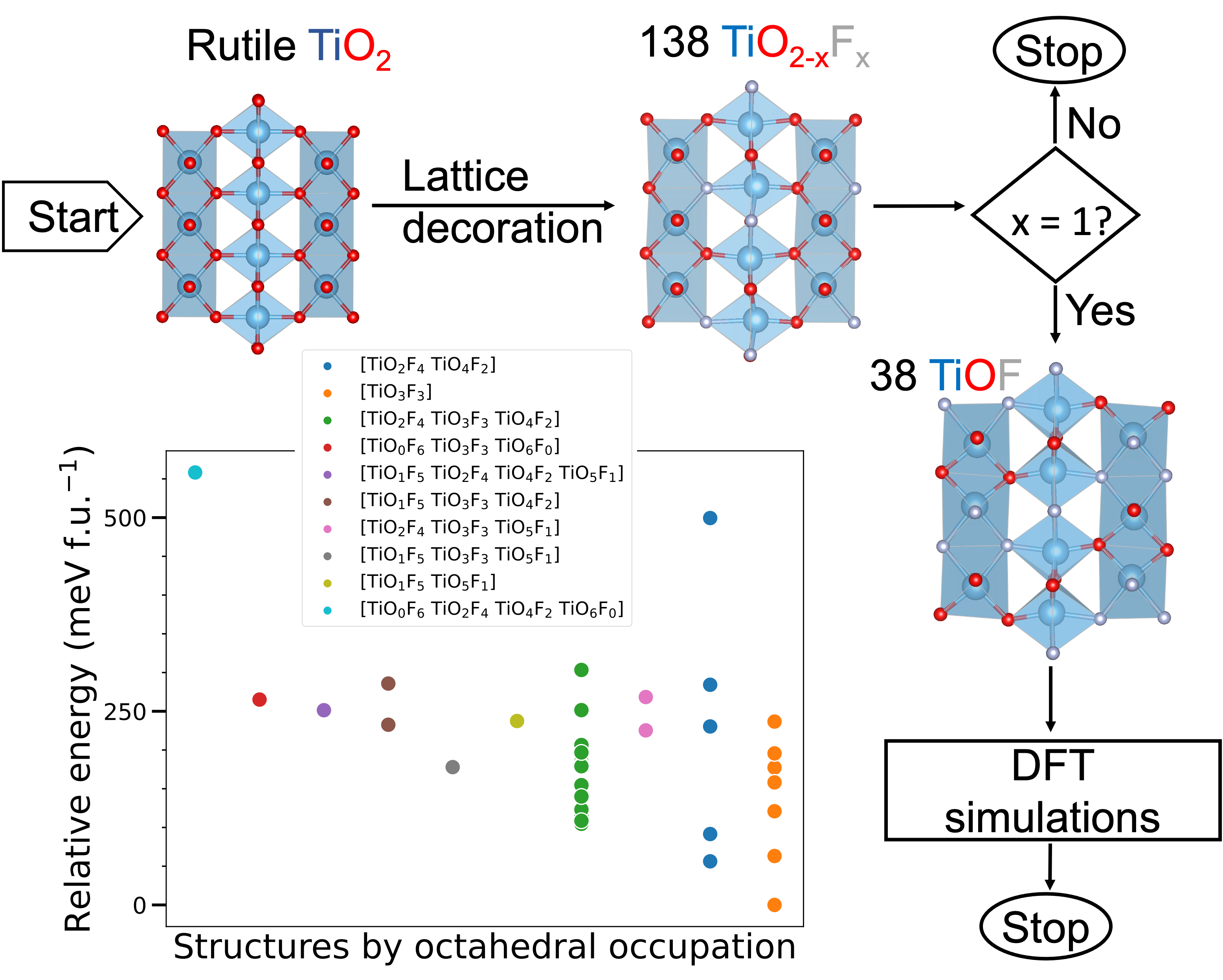}
    \caption{Workflow used to find the ground state anion ordering in TiOF, along with the relative energies of each of the 38 structures grouped according to unique TiO$_x$F$_{6-x}$ octahedral environments.}
    \label{fig1}
\end{figure*}

Here, we assess the effect of fluoride substitution and the resulting anion ordering on the interplay between Peierls and Mott physics in rutile structured heteroanionic compounds. Using density functional theory (DFT) calculations, we examine the crystal and electronic structure of TiOF, a  Ti$^{3+}$ ($3d^1$ electronic configuration) analogue of VO$_2$ \cite{chamberland1967preparation} containing [TiO$_3$F$_3$] octahedral units. TiOF has been synthesized in the rutile structure, but no evidence of long-range anion ordering was found \cite{cumby2018high}. 
Diffuse X-Ray scattering measurements, however, revealed extended correlations along $\langle$110$\rangle$ planes that could be a consequence of covalency-driven anion ordering.  
To study the effect of local anion ordering on the electronic structure of TiOF, we formulate an anion-ordered model of TiOF. Our findings reveal that TiOF exhibits a Peierls-like distortion that drives the anion ordering along the $\langle$110$\rangle$ planes. As electronic correlation increases, TiOF undergoes two first-order phase transitions: first, transitioning from a nonmagnetic insulator to a ferromagnetic half-metal, and then progressing to become a ferromagnetic  insulator, characterized by the absence of the Peierls-like distortion. We further find that the electronic configuration of the transition metal plays an important role in  determining the propensity to and pattern of anion order in rutile heteroanionic materials.

\section{Methods}
Our workflow to find the equilibrium ground state structures is illustrated in \autoref{fig1}. To begin, we used the ideal rutile TiO$_2$ structure ($P4_2/mnm$) as the prototype structure from which candidate TiOF structures were enumerated by decorating O sites with F atoms. The structure enumeration code ENUMLIB \cite{hart2008algorithm}, as implemented in the integrated cluster expansion toolkit (ICET) package \cite{aangqvist2019icet} was used for this purpose. We generated all potential TiO$_{2-x}$F$_x$ ($0\leq x\leq 1$) structures compatible with $1\times1\times2$, $1\times2\times1$, and $2\times1\times1$ supercells of the rutile basis vectors and extracted those with the TiOF stoichiometry. We identified 38 symmetrically unique TiOF structures. 

We then performed structural relaxations of the ion and cell coordinates for each of the 38 candidate structures using DFT simulations as described below. To ensure we identified the correct ground state, the 38 structures were relaxed (volume and atomic positions) with the spins on the Ti atoms initialized with ferromagnetic and antiferromagnetic spin configurations. Non-spinpolarized calculations were also performed. The ground state of each anion configuration was determined by identifying the structure with the lowest total energy amongst all spin-orderings. The results of this process allowed us to obtain the energies of all 38 structures (\autoref{fig1}), grouped according to unique [TiO$_{x}$F$_{6-x}$] octahedral environments, relative to the lowest energy configuration with exclusive TiO$_3$F$_3$ octahedra.

\begin{figure}
\centering
\includegraphics[width=0.49\textwidth]{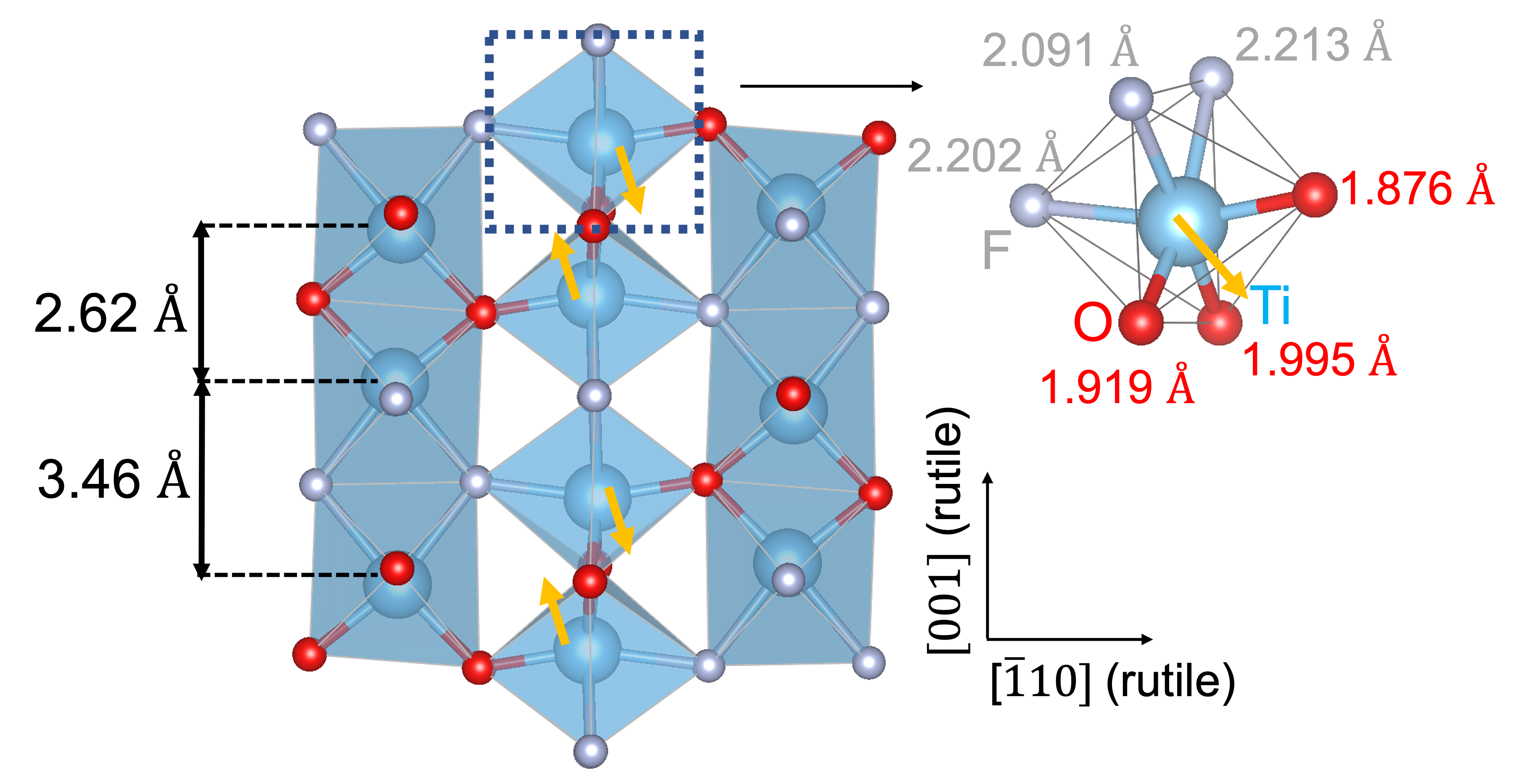}
    \caption{(a) Atomic structure of the $P2_1/c$ monoclinic phase of TiOF. The inset shows the \textit{fac} ordered TiO$_3$F$_3$ octahedra. The Ti off-centering is highlighted by the yellow arrows.}
    \label{fig2}
\end{figure}

DFT simulations were performed with the Vienna Ab-Initio Simulation Package (VASP) code \cite{kresse1996efficient}. We used the projector augmented wave (PAW) technique \cite{kresse1999ultrasoft} with a plane-wave energy cutoff of 650\,eV and the Perdew, Burke, and Ernzerhof (PBE) generalized gradient (GGA)  exchange-correlation functional \cite{perdew1996generalized}. All calculations were performed using dense well-converged $\Gamma$-centered $k$-meshes of at least 4,000 $k$-points per reciprocal atom, and each structure was relaxed until the forces on the atoms were less than $5\times 10^{-4}$\,eV \AA$^{-1}$. The Atomic Simulation Environment (ASE) was used to aid all calculations and post-processing \cite{larsen2017atomic}.

\section{Results and Discussion}
\subsection{Atomic Structure}
The ground state structure of the anion-ordered TiOF compound  exhibits monoclinic $P2_1/c$ symmetry (\autoref{fig2}), which is enforced by the arrangement of \textit{fac}-[TiO$_3$F$_3$] ordered octahedra. The structure shows twisted Ti-Ti dimers along the [001] rutile direction with an intradimer distance of 2.62\,\AA. Although the monoclinic TiOF phase shares similarities with the monoclinic phases of VO$_2$ and NbO$_2$ \cite{eyert2002metal,eyert2002metal2}, it exhibits key differences. First, the ordered \textit{fac}-[TiO$_3$F$_3$] octahedra cause a symmetry reduction that prohibits TiOF from exhibiting the ideal rutile symmetry. Second, the Ti-Ti dimerization is globally stable, confirmed with our lattice dynamical simulations shown in the Supplemental Materials (SM) \cite{Supp}, whereas the $P2_1/c$ structure without Ti-Ti dimers (uniform Ti-Ti distances) is spontaneously unstable.

We attribute the origin of the stability of the $P2_1/c$ phase to the \textit{fac}-configuration, which is characterized by [TiO$_3$F$_3$] octahedra having alternating faces containing exclusively O or F anions (\autoref{fig2}).This anion ordering within the heteroleptic unit leads to the displacement of Ti ions towards the octahedral face containing O atoms, as indicated by the difference between Ti-O  (average bond length of 1.93\,\AA) and Ti-F bonds (2.17\,\AA) length.
As a result, the Ti displacements exhibit two distinct components as indicated in \autoref{fig2}. The first component involves a shift along the [001] rutile direction, marked by Ti ions moving out of the center toward the O-O edge of each octahedron. This displacement facilitates the creation of Ti-Ti dimers and the ensuing metal-metal bonding (\emph{vide infra}). The second component, represented by antiparallel displacements perpendicular to the [001] rutile direction, induces the twisting of the Ti-Ti dimers.

\subsection{Electronic Structure}
\begin{figure}
    \centering
    \includegraphics[width=0.49\textwidth]{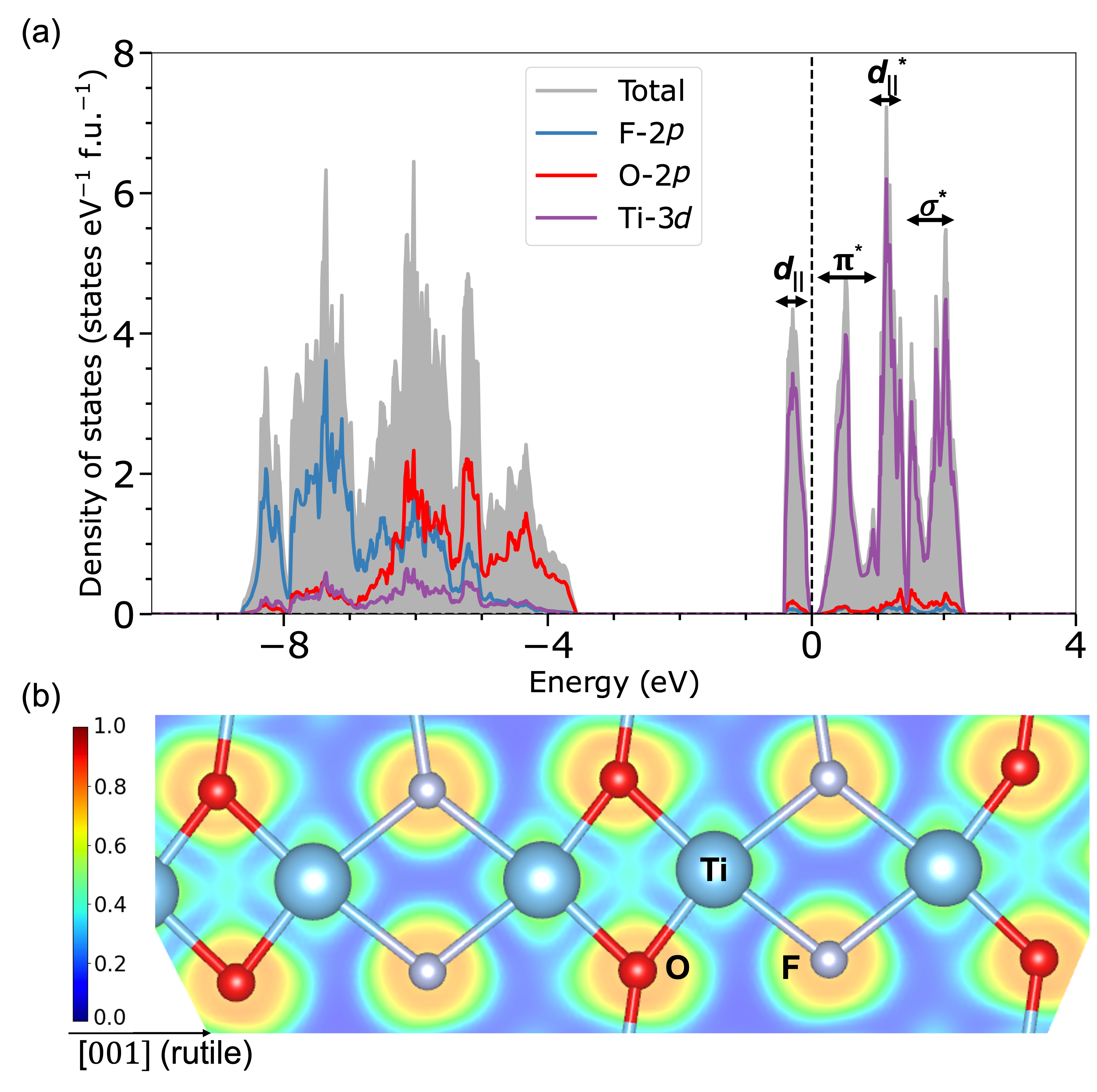}
    \caption{(a) Atom and orbital resolved density of states of  $P2_1/c$  TiOF. (b) Electronic localization function (ELF) projected along the (110) plane. Ti-Ti dimers are visible along the [001] rutile direction where the ELF shows mixed  semicovalent   metallic bonding.}   \label{fig3}
\end{figure}

\autoref{fig3}a shows the electronic density of states (DOS) of TiOF. The compound is insulating with a small gap separating Ti-derived states with narrow bandwidths. 
The formation of the Ti-Ti dimer leads to the presence of an occupied metal-metal bonding $d_{||}$ ($S=0$ singlet) state and an unoccupied anti-bonding $d^*_{||}$ state derived from the  $d_{xy}$ orbitals. The  conduction band minimum is  8\,meV (at the DFT-PBE level) above the  $d_{||}$ state and comprises anti-bonding $d^*_{\pi}$ states  formed from Ti $d_{xz}/d_{yz}$ orbitals.
The O and F $2p$ states are primarily distributed over the energy range of -9.0 to -3.5\,eV with some degree of hybridization observed between the O $2p$ states and the Ti $d$ states. 
These electronic structure features make TiOF remarkably similar to the monoclinic phase of VO$_2$ \cite{eyert2002metal} and some $d^1$ trirutile compounds \cite{Miller_2022,Schueller_2021}. 

\autoref{fig3}b shows the electron localization function (ELF) projected on the (110) plane \cite{silvi1994classification}. 
ELF values greater than 0.7 are typically used to identify regions of strong localization due to covalent bond formation or electron lone
pairs. ELF values between ranging from 0.7 to 0.2 are indicative of metallic bonding, i.e., the degree of localization for a homogeneous elecron gas \cite{Koumpouras2020}.
Here we find the ELF ranges from 0.4 to 0.5 between adjacent Ti atoms bridged by oxide anions, supporting our identification of  Ti-Ti dimer formation and mixed covalent and metallic bonding defining the low-energy electronic structure formed by the $d_{||}$ and $d^*_{\pi}$ states.
In addition, the shorter Ti-O bonds are more covalent compared to the longer more ionic Ti-F bonds, which can be deduced by the shape of the ELF localized on the anion site and its radially extent along the bonds towards Ti. 
Examining the ELF of the monoclinic phase of VO$_2$, we see a similar type of metal-metal bond formation and V-O hybridization \cite{Supp}.

\subsection{Role of Electronic Correlation}
\begin{figure}
    \centering
\includegraphics[width=0.48\textwidth]{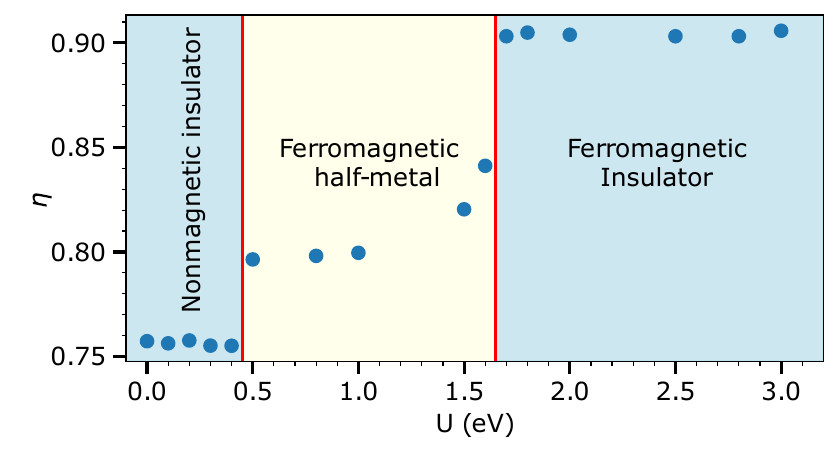}
    \caption{TiOF phase map showing the dependence of $\eta$, defined as the ratio of the  short to long Ti-Ti distances, as a function of the Ti $d$ onsite Coulomb interaction $U$. The electron-electron interactions weaken the dimerization ($\eta$ increases), by preferentially localizing the electrons on the Ti sites and disfavoring  metal-metal bonding.}
    \label{fig4}
\end{figure}

To assess the influence of electron-electron interactions on TiOF, we also performed  atomic relaxation of the ground state $P2_1/c$ structure using the DFT plus Hubbard $U$ method. We used the Dudarev et al.\ approach \cite{dudarev1998electron} with a single effective $U_\mathrm{eff} = U-J$, hereafter $U$, ranging from 0 to 3\,eV on the Ti $3d$ manifold. \autoref{fig4} shows that the onsite interactions weaken the strength of the Ti-Ti dimer. We quantify this reduction in the dimer strength structurally by using the paramater $\eta$, which we define as the ratio of the equilibrium short and long Ti-Ti bond distances for a given $U$ value. 

With increasing $U$,  $\eta$ increases such that the short- and long Ti-Ti distances become more similar. In addition, we find, two abrupt changes in $\eta$, indicative of two first-order phases transitions associated with changes in the electronic and magnetic structure. 
For $0.0 \le U \le 0.4$\,eV, TiOF exhibits nonmagnetic insulating behavior consistent with the aforementioned electronic structure description. 
As $U$ increases, however, we find that the singlet state does not form for $0.5\le U \le 1.6$,eV, and a ferromagnetic half-metallic state is stable.
This state is unlikely to be long-ranged ordered with a well-defined Curie temperature,  but rather represents an $S=1/2$ paramagnetic state of TiOF, which is not simple to capture in our simulation cell and requires spin disordered polymorphous configurations at the DFT level \cite{PhysRevB.97.035107}.
Finally for $1.6 \le U \le 3$\,eV, TiOF transforms into a ferromagnetic insulator, where both spin-channels are gapped. In a similar way this phase is likely better described as a paramagnetic Mott insulator with 1 electron per Ti localized on the Ti site owing to the strong Coulomb interactions.

\subsection{Role of Anion Ordering}
\begin{figure}
    \centering
    \includegraphics[width=0.49\textwidth]{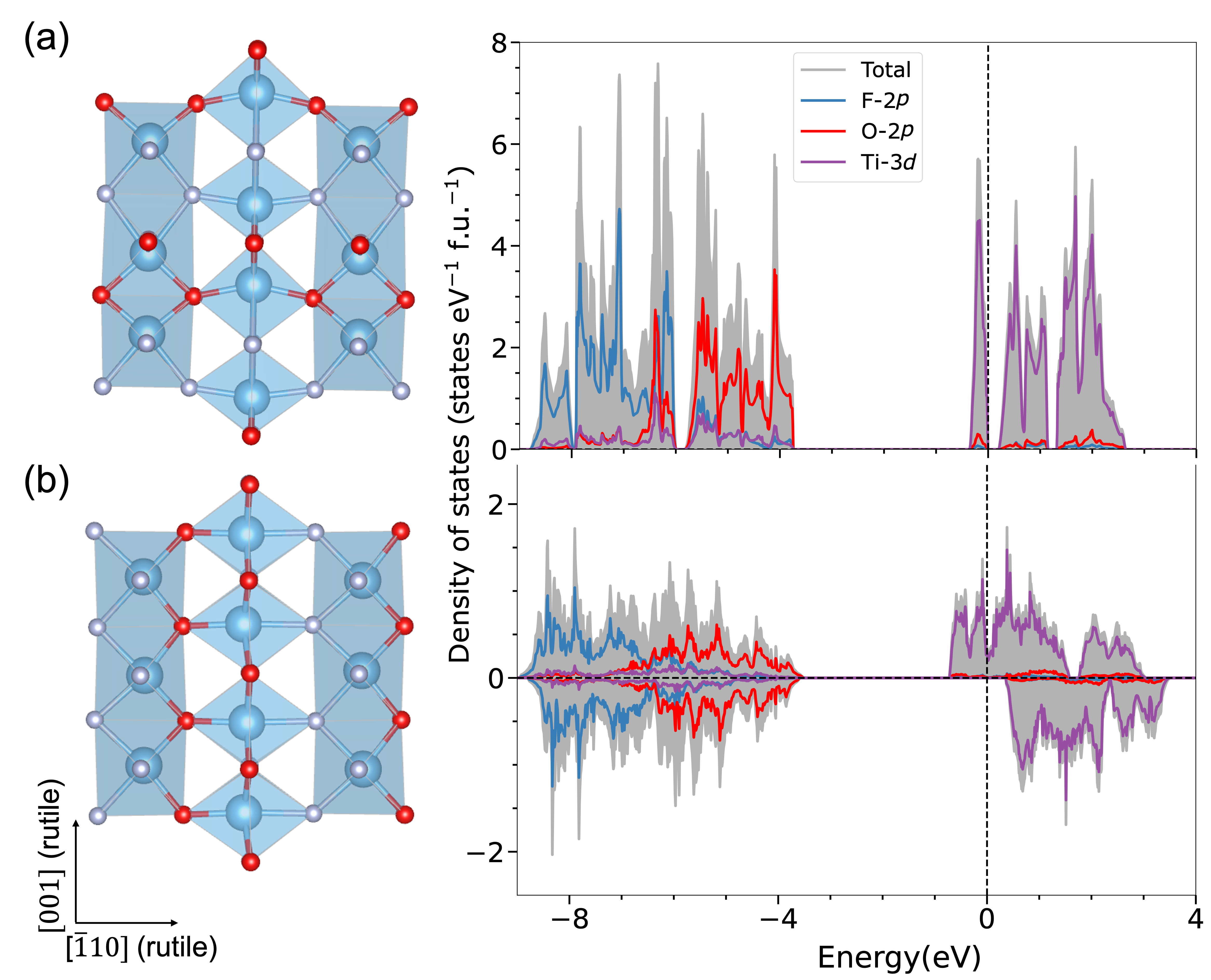}
    \caption{Atomic structure, viewed along the [110] direction, and the electronic density of states of the (a) $P\bar{4}2_{1}m$ and (b) $Cc$ phase. The shared edges comprising O-O and F-F pairs in (a) facilitate the formation of an insulating singlet state whereas the mixed anion O-F pairs in (b) give rise to a ferromagnetic half-metallic state.}
    \label{fig5}
\end{figure}

In oxyfluorides, \textit{fac} ordered [MO$_3$F$_3$] heteroleptic units are frequently encountered \cite{harada2019heteroanionic}. Although the ground state $P2_1/c$ structure of TiOF adheres to this pattern, we seek to understand how anion ordering influences the stability of TiOF and its role in the formation of Ti-Ti dimers.
To that end, we examine two of the atomic structures obtained through our structural search of TiOF in more detail. They exhibit different anion orderings: The first structure is characterized by tetragonal $P\bar{4}2_{1}m$ symmetry and consists of alternating \textit{cis}-[TiO$_4$F$_2$] and \textit{cis}-[TiO$_2$F$_4$] octahedra along the [001] rutile direction (\autoref{fig5}a). This structure shows the same edge-shared octahedral connectivity as found in the monoclinic \textit{P2$_1$/c} phase (\autoref{fig3}), i.e., , where O-O and F-F pairs comprise the shared edge.  The second structure exhibits ordered \textit{fac}-[TiO$_3$F$_3$] heteroleptic units and possesses orthorhombic polar $Pmn2_1$ symmetry. It is distinct from the ground state centrosymmetric $P2_1/c$ phase, however, as it features edge-shared octahedra with the shared edge  comprising dissimliar anions, i.e., O-F pairs  (\autoref{fig5}b). 

Following structural relaxation of these phases, we find that the tetragonal $P\bar{4}2_{1}m$ polymorph exhibits spontaneous Ti-Ti dimerization, with a Ti-Ti dimer distance of 2.56\,\AA\ and out-of-center displacements. In contrast, the  orthorhombic $Pmn2_1$ phase does not display dimerization, although the Ti cations do displace as before towards the oxide anions. In this structure, the change in the oxide anion ordering leads to Ti displacements that are mainly transverse to the 1D octhaedral chain [001] direction.
The tetragonal $P\bar{4}2_{1}m$ phase is a nonmagnetic insulator, exhibiting an electronic band gap of 0.2\,eV (\autoref{fig5}a). Despite being energetically 56\,meV per formula unit (f.u.) higher in energy compared to the ground state structure, it is dynamically stable, as confirmed by our phonon calculations \cite{Supp}. Interestingly, we find that the anion ordering with homogeneous anion pairs comprising the shared octahedral edges drives the Ti-Ti dimerization despite the Ti cations formally exhibiting mixed valence  (Ti$^{3+\delta}$/Ti$^{3-\delta}$) locally imposed by the heteroleptic [TiO$_4$F$_2$]/[TiO$_2$F$_4$] units in the tetragonal structure.

The orthorhombic \textit{Pmn2$_1$} phase is a half-metal, with Ti atoms exhibiting a local magnetic moment of $\sim0.9$\,$\mu_B$. 
This phase is $\sim128$\,meV/f.u.\  higher in energy with respect to the monoclinic $P2_1/c$ structure. Unlike the tetragonal $P\bar{4}2_{1}m$ phase, hoever, it is dynamically unstable as shown in the SM \cite{Supp}. To find an equilibrium configruation, we systematically ‘froze-in’ linear combinations of the unstable modes, and performed structural optimization. This procedure led to a structure with the monoclinic $Cc$ space group, which is higher in energy by 115\,meV/f.u. compared to the ground state structure. 
The monoclinic $Cc$ phase, despite displaying half-metallic behavior (\autoref{fig5}b) with a magnetic moment of $\sim0.9$\,$\mu_B$/Ti, exhibits Ti-Ti dimerization. Interestingly, the intradimer distance is 2.84\,\AA, which is larger than that observed in the $P2_1/c$ and $P\bar{4}2_{1}m$ phases. This increased Ti-Ti dimer distance prohibits the formation of a nonmagnetic singlet state, resulting in metallicity in the majority spin channel.

Our analysis indicates that TiOF undegoes a Peierls distortion, regardless of the anion ordering. The conseqeunce of the distortion on the electronic structure, however, the depends on the anionic character between edge-shared octahedra. When octahedra share O-O/F-F edges, they promote greater dimerization, i.e., shorter Ti-Ti distances, and subsequently the formation of a singlet state. This leads to the opening of a band gap and a decrease in total energy. Conversely, when octahedra share mixed anion O-F/O-F edges, there is competition between the Coulombic interactions arising from the ionic Ti-F bond and the formation of Ti-Ti dimers with metal-metal bonding. This results in weaker dimerization, reduced metal-metal bonding,  and the ferromagnetic half-metallic state.
By comparing the monoclinic $P2_1/c$ ground state and the metastable tetragonal $P\bar{4}2_1m$ phase, we can conclude that ordered \textit{fac}-[TiO$_3$F$_3$] heteroleptic units are more favorable than \textit{cis}-TiO$_4$F$_2$] and \textit{cis}-[TiO$_2$F$_4$] octahedra. All these observations suggest that although TiOF exhibits a preference for locally ordered \textit{fac}-TiO$_3$F$_3$] octahedra, the Peierls-like distortion ultimately dictates their assembly to have homogeneous anion pairs (O-O and F-F) comprising the shared edges.

\subsection{Role of Cation Configuration}

\begin{figure}
    \centering
    \includegraphics[width=0.48\textwidth]{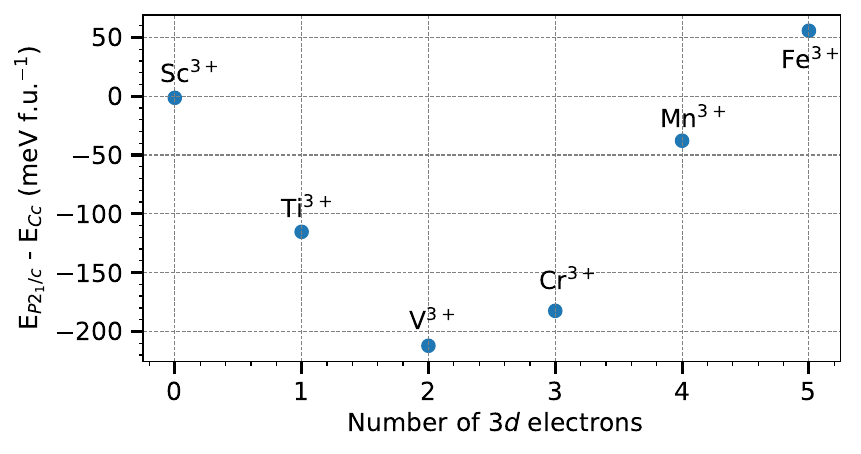}
    \caption{Relative energies of the $Cc$ and the $P2_1/c$ polymorphs of $M$OF ($M =$ Sc, Ti, V, Cr, Mn, and Fe). A change in stability between the preferred \textit{fac}-$M$O$_3$F$_3$ ordering occurs between Mn to Fe.}
    \label{fig6}
\end{figure}

To investigate the influence of the electronic configuration of the transition metal cation on the prefered anion ordering in  rutile-structured oxyfluorides $M$OF ($M=$ Sc, Ti, V, Cr, Mn, and Fe), we carried out the following computational experiment. We substitute Ti with the other $d$ transition metals in TiOF within the $P2_1/c$ and $Cc$ polymorphs and then perform a full atomic and volumetric relaxation within these symmetries. 
\autoref{fig6}  shows the relative energy difference between the $M$OF $P2_1/c$ phase, characterized by an O-O/F-F edge-sharing arrangement (\autoref{fig2}), and the monoclinic $Cc$  phase, exhibiting octahedra that share edges with O-F/O-F arrangement(\autoref{fig5}b).

For $M=$ Sc with a $d^0$ electronic configuration, we find no specific ordering preference as the relative energy difference is essentially zero. For electronic configurations spanning from $d^1$ to $d^3$ ($M=$ Ti, V, and Cr), however, we find a strong preference for the O-O/F-F edge-sharing configuration with  $P2_1/c$ symmetry. The competition between the two polymorphs  increases with increasing $3d$ electron filling, e.g., moving from $d^3$ to $d^4$ ($M=$ Mn). Ultimately, the O-F/O-F edge-sharing arrangement with $Cc$ symmetry is favored for the $d^5$ electronic configuration ($M=$ Fe).

We note that although including the effects of electronic correlation and magnetism may change the absolute energy scales for the energy differences, we expect a similar trend as the octahedral assembly depends on the metal-metal bonding tuned by the electron filling.
We arrive at this conclusion and  understand the trend shown in \autoref{fig6}  using arguments that describe the stability of metal dimers in homonanionic rutile-structured oxides.
The strength of the metal-metal bond increases as the electronic configuration changes from $d^1$ to $d^3$ \cite{hiroi2015structural}. This characteristic is reflected in the preference for the O-O/F-F edge-sharing arrangement in MOF ($M =$Ti, V, Cr). As the $d$-electron filling increases, the $d$ orbitals contract, decreasing the orbital overlap and weakening the metal-metal bonding. This, in turn, leads to enhanced competition between the two anion ordering arrangements in MnOF and the preference for the O-F/O-F edge-sharing arrangement in FeOF. 
In the case of a $d^0$ electronic configuration, the metal-metal bond is formally inactive, leading to no specific preference for any particular anion ordering.

\section{Conclusions}
 We performed electronic structure calculations to find that the anion order in TiOF is driven by the competition of the ionic nature of the Ti-F bond with the tendency of the Ti$^{3+}$ cation to dimerize and form a singlet within the edge-shared rutile framework. In particular, we examined the importance of O/F anion ordering within a single heteroleptic octahedron as well as the longer-range anion order along the rutile chains. We found  that when the anion pairs of the same type (O-O or F-F) comprise the shared octahedral edges, Ti-Ti dimerization and a singlet state occur. 
 In addition, we found that including the effects of strong electron-electron interactions led to a competition with the Peierls distortion by suppressing the formation of a Ti-Ti metal-metal bond. By increasing the correlation strength, we uncovered two first-order phase transitions: first from a nonmagnetic insulator to a ferromagnetic half-metal and then to a ferromagnetic insulator.   Although ferromagnetically ordered in our simulations, these correlation-stabilized states are likely to be paramagnetic. 
Furthermore, we  established that the electronic configuraiton of the transition metal cation plays a decisive role in dictating the long-range anion ordering in oxyfluorides. Future work will focus on understanding and uncovering trends in the relation between anion ordering and cationic behavior at finite temperatures, with an aim to design heteronanionic materials that exhibit collective phenomena through cation sublattice control.

\begin{acknowledgments}
This work was funded by the National Science Foundation (NSF) under award number DMR-2011208. The authors are grateful to Prof.\ James Cumby for discussions on published and unpublished experimental TiOF data, and Dr.\ Jaye Harada for contributions to exploratory  computational studies of TiOF. DFT simulations were performed using the Quest HPC facility at Northwestern, Bridges-2 at the Pittsburgh Supercomputing Center from the Advanced Cyberinfrastructure Coordination Ecosystem: Services and Support (ACCESS) program, which is supported by NSF grants 2138259, 2138286, 2138307, 2137603, and 2138296; and the Carbon cluster at the Center for Nanoscale Materials, a U.S.\ Department of Energy Office of Science User Facility, supported by the U.S.\ DOE, Office of Basic Energy Sciences, under Contract No.\ DE-AC02-06CH11357.

\end{acknowledgments}
\bibliography{references}

\begin{thebibliography}{41}%
\makeatletter
\providecommand \@ifxundefined [1]{%
 \@ifx{#1\undefined}
}%
\providecommand \@ifnum [1]{%
 \ifnum #1\expandafter \@firstoftwo
 \else \expandafter \@secondoftwo
 \fi
}%
\providecommand \@ifx [1]{%
 \ifx #1\expandafter \@firstoftwo
 \else \expandafter \@secondoftwo
 \fi
}%
\providecommand \natexlab [1]{#1}%
\providecommand \enquote  [1]{``#1''}%
\providecommand \bibnamefont  [1]{#1}%
\providecommand \bibfnamefont [1]{#1}%
\providecommand \citenamefont [1]{#1}%
\providecommand \href@noop [0]{\@secondoftwo}%
\providecommand \href [0]{\begingroup \@sanitize@url \@href}%
\providecommand \@href[1]{\@@startlink{#1}\@@href}%
\providecommand \@@href[1]{\endgroup#1\@@endlink}%
\providecommand \@sanitize@url [0]{\catcode `\\12\catcode `\$12\catcode
  `\&12\catcode `\#12\catcode `\^12\catcode `\_12\catcode `\%12\relax}%
\providecommand \@@startlink[1]{}%
\providecommand \@@endlink[0]{}%
\providecommand \url  [0]{\begingroup\@sanitize@url \@url }%
\providecommand \@url [1]{\endgroup\@href {#1}{\urlprefix }}%
\providecommand \urlprefix  [0]{URL }%
\providecommand \Eprint [0]{\href }%
\providecommand \doibase [0]{https://doi.org/}%
\providecommand \selectlanguage [0]{\@gobble}%
\providecommand \bibinfo  [0]{\@secondoftwo}%
\providecommand \bibfield  [0]{\@secondoftwo}%
\providecommand \translation [1]{[#1]}%
\providecommand \BibitemOpen [0]{}%
\providecommand \bibitemStop [0]{}%
\providecommand \bibitemNoStop [0]{.\EOS\space}%
\providecommand \EOS [0]{\spacefactor3000\relax}%
\providecommand \BibitemShut  [1]{\csname bibitem#1\endcsname}%
\let\auto@bib@innerbib\@empty
\bibitem [{\citenamefont {Pandey}(2010)}]{pandey2010correlation}%
  \BibitemOpen
  \bibfield  {author} {\bibinfo {author} {\bibfnamefont {S.~K.}\ \bibnamefont
  {Pandey}},\ }\bibfield  {title} {\bibinfo {title} {{Correlation induced
  half-metallicity in a ferromagnetic single-layered compound:
  Sr$_2$CoO$_4$}},\ }\href@noop {} {\bibfield  {journal} {\bibinfo  {journal}
  {Physical Review B}\ }\textbf {\bibinfo {volume} {81}},\ \bibinfo {pages}
  {035114} (\bibinfo {year} {2010})}\BibitemShut {NoStop}%
\bibitem [{\citenamefont {Monthoux}\ \emph {et~al.}(1991)\citenamefont
  {Monthoux}, \citenamefont {Balatsky},\ and\ \citenamefont
  {Pines}}]{monthoux1991toward}%
  \BibitemOpen
  \bibfield  {author} {\bibinfo {author} {\bibfnamefont {P.}~\bibnamefont
  {Monthoux}}, \bibinfo {author} {\bibfnamefont {A.}~\bibnamefont {Balatsky}},\
  and\ \bibinfo {author} {\bibfnamefont {D.}~\bibnamefont {Pines}},\ }\bibfield
   {title} {\bibinfo {title} {Toward a theory of high-temperature
  superconductivity in the antiferromagnetically correlated cuprate oxides},\
  }\href@noop {} {\bibfield  {journal} {\bibinfo  {journal} {Physical Review
  Letters}\ }\textbf {\bibinfo {volume} {67}},\ \bibinfo {pages} {3448}
  (\bibinfo {year} {1991})}\BibitemShut {NoStop}%
\bibitem [{\citenamefont {Moshnyaga}\ \emph {et~al.}(2009)\citenamefont
  {Moshnyaga}, \citenamefont {Gehrke}, \citenamefont {Lebedev}, \citenamefont
  {Sudheendra}, \citenamefont {Belenchuk}, \citenamefont {Raabe}, \citenamefont
  {Shapoval}, \citenamefont {Verbeeck}, \citenamefont {Van~Tendeloo},\ and\
  \citenamefont {Samwer}}]{moshnyaga2009electrical}%
  \BibitemOpen
  \bibfield  {author} {\bibinfo {author} {\bibfnamefont {V.}~\bibnamefont
  {Moshnyaga}}, \bibinfo {author} {\bibfnamefont {K.}~\bibnamefont {Gehrke}},
  \bibinfo {author} {\bibfnamefont {O.~I.}\ \bibnamefont {Lebedev}}, \bibinfo
  {author} {\bibfnamefont {L.}~\bibnamefont {Sudheendra}}, \bibinfo {author}
  {\bibfnamefont {A.}~\bibnamefont {Belenchuk}}, \bibinfo {author}
  {\bibfnamefont {S.}~\bibnamefont {Raabe}}, \bibinfo {author} {\bibfnamefont
  {O.}~\bibnamefont {Shapoval}}, \bibinfo {author} {\bibfnamefont
  {J.}~\bibnamefont {Verbeeck}}, \bibinfo {author} {\bibfnamefont
  {G.}~\bibnamefont {Van~Tendeloo}},\ and\ \bibinfo {author} {\bibfnamefont
  {K.}~\bibnamefont {Samwer}},\ }\bibfield  {title} {\bibinfo {title}
  {Electrical nonlinearity in colossal magnetoresistance manganite films:
  Relevance of correlated polarons},\ }\href@noop {} {\bibfield  {journal}
  {\bibinfo  {journal} {Physical Review B}\ }\textbf {\bibinfo {volume} {79}},\
  \bibinfo {pages} {134413} (\bibinfo {year} {2009})}\BibitemShut {NoStop}%
\bibitem [{\citenamefont {Rondinelli}\ \emph {et~al.}(2008)\citenamefont
  {Rondinelli}, \citenamefont {Caffrey}, \citenamefont {Sanvito},\ and\
  \citenamefont {Spaldin}}]{PhysRevB.78.155107}%
  \BibitemOpen
  \bibfield  {author} {\bibinfo {author} {\bibfnamefont {J.~M.}\ \bibnamefont
  {Rondinelli}}, \bibinfo {author} {\bibfnamefont {N.~M.}\ \bibnamefont
  {Caffrey}}, \bibinfo {author} {\bibfnamefont {S.}~\bibnamefont {Sanvito}},\
  and\ \bibinfo {author} {\bibfnamefont {N.~A.}\ \bibnamefont {Spaldin}},\
  }\bibfield  {title} {\bibinfo {title} {{Electronic properties of bulk and
  thin film SrRuO$_3$: Search for the metal-insulator transition}},\
  }\href@noop {} {\bibfield  {journal} {\bibinfo  {journal} {Physical Review
  B}\ }\textbf {\bibinfo {volume} {78}},\ \bibinfo {pages} {155107} (\bibinfo
  {year} {2008})}\BibitemShut {NoStop}%
\bibitem [{\citenamefont {Puggioni}\ \emph {et~al.}(2018)\citenamefont
  {Puggioni}, \citenamefont {Stroppa},\ and\ \citenamefont
  {Rondinelli}}]{puggioni2018design}%
  \BibitemOpen
  \bibfield  {author} {\bibinfo {author} {\bibfnamefont {D.}~\bibnamefont
  {Puggioni}}, \bibinfo {author} {\bibfnamefont {A.}~\bibnamefont {Stroppa}},\
  and\ \bibinfo {author} {\bibfnamefont {J.~M.}\ \bibnamefont {Rondinelli}},\
  }\bibfield  {title} {\bibinfo {title} {Design of a polar half-metallic
  ferromagnet with accessible and enhanced electric polarization},\ }\href@noop
  {} {\bibfield  {journal} {\bibinfo  {journal} {Physical Review Materials}\
  }\textbf {\bibinfo {volume} {2}},\ \bibinfo {pages} {114403} (\bibinfo {year}
  {2018})}\BibitemShut {NoStop}%
\bibitem [{\citenamefont {Morin}(1959)}]{PhysRevLett.3.34}%
  \BibitemOpen
  \bibfield  {author} {\bibinfo {author} {\bibfnamefont {F.~J.}\ \bibnamefont
  {Morin}},\ }\bibfield  {title} {\bibinfo {title} {Oxides which show a
  metal-to-insulator transition at the neel temperature},\ }\href
  {https://doi.org/10.1103/PhysRevLett.3.34} {\bibfield  {journal} {\bibinfo
  {journal} {Phys. Rev. Lett.}\ }\textbf {\bibinfo {volume} {3}},\ \bibinfo
  {pages} {34} (\bibinfo {year} {1959})}\BibitemShut {NoStop}%
\bibitem [{\citenamefont {Woodley}(2008)}]{woodley2008mechanism}%
  \BibitemOpen
  \bibfield  {author} {\bibinfo {author} {\bibfnamefont {S.~M.}\ \bibnamefont
  {Woodley}},\ }\bibfield  {title} {\bibinfo {title} {The mechanism of the
  displacive phase transition in vanadium dioxide},\ }\href@noop {} {\bibfield
  {journal} {\bibinfo  {journal} {Chemical Physics Letters}\ }\textbf {\bibinfo
  {volume} {453}},\ \bibinfo {pages} {167} (\bibinfo {year}
  {2008})}\BibitemShut {NoStop}%
\bibitem [{\citenamefont {Stefanovich}\ \emph {et~al.}(2000)\citenamefont
  {Stefanovich}, \citenamefont {Pergament},\ and\ \citenamefont
  {Stefanovich}}]{stefanovich2000electrical}%
  \BibitemOpen
  \bibfield  {author} {\bibinfo {author} {\bibfnamefont {G.}~\bibnamefont
  {Stefanovich}}, \bibinfo {author} {\bibfnamefont {A.}~\bibnamefont
  {Pergament}},\ and\ \bibinfo {author} {\bibfnamefont {D.}~\bibnamefont
  {Stefanovich}},\ }\bibfield  {title} {\bibinfo {title} {{Electrical switching
  and Mott transition in VO$_2$}},\ }\href@noop {} {\bibfield  {journal}
  {\bibinfo  {journal} {Journal of Physics: Condensed Matter}\ }\textbf
  {\bibinfo {volume} {12}},\ \bibinfo {pages} {8837} (\bibinfo {year}
  {2000})}\BibitemShut {NoStop}%
\bibitem [{\citenamefont {Kim}\ \emph {et~al.}(2013)\citenamefont {Kim},
  \citenamefont {Kim}, \citenamefont {Kang},\ and\ \citenamefont
  {Min}}]{kim2013correlation}%
  \BibitemOpen
  \bibfield  {author} {\bibinfo {author} {\bibfnamefont {S.}~\bibnamefont
  {Kim}}, \bibinfo {author} {\bibfnamefont {K.}~\bibnamefont {Kim}}, \bibinfo
  {author} {\bibfnamefont {C.-J.}\ \bibnamefont {Kang}},\ and\ \bibinfo
  {author} {\bibfnamefont {B.}~\bibnamefont {Min}},\ }\bibfield  {title}
  {\bibinfo {title} {{Correlation-assisted phonon softening and the
  orbital-selective Peierls transition in VO$_2$}},\ }\href@noop {} {\bibfield
  {journal} {\bibinfo  {journal} {Physical Review B}\ }\textbf {\bibinfo
  {volume} {87}},\ \bibinfo {pages} {195106} (\bibinfo {year}
  {2013})}\BibitemShut {NoStop}%
\bibitem [{\citenamefont {Szymanski}\ \emph {et~al.}(2019)\citenamefont
  {Szymanski}, \citenamefont {Walters}, \citenamefont {Puggioni},\ and\
  \citenamefont {Rondinelli}}]{szymanski2019design}%
  \BibitemOpen
  \bibfield  {author} {\bibinfo {author} {\bibfnamefont {N.~J.}\ \bibnamefont
  {Szymanski}}, \bibinfo {author} {\bibfnamefont {L.~N.}\ \bibnamefont
  {Walters}}, \bibinfo {author} {\bibfnamefont {D.}~\bibnamefont {Puggioni}},\
  and\ \bibinfo {author} {\bibfnamefont {J.~M.}\ \bibnamefont {Rondinelli}},\
  }\bibfield  {title} {\bibinfo {title} {{Design of heteroanionic MoON
  exhibiting a peierls metal-insulator transition}},\ }\href@noop {} {\bibfield
   {journal} {\bibinfo  {journal} {Physical Review Letters}\ }\textbf {\bibinfo
  {volume} {123}},\ \bibinfo {pages} {236402} (\bibinfo {year}
  {2019})}\BibitemShut {NoStop}%
\bibitem [{\citenamefont {Hartman}\ \emph {et~al.}(2018)\citenamefont
  {Hartman}, \citenamefont {Cho},\ and\ \citenamefont
  {Mishra}}]{hartman2018multiferroism}%
  \BibitemOpen
  \bibfield  {author} {\bibinfo {author} {\bibfnamefont {S.~T.}\ \bibnamefont
  {Hartman}}, \bibinfo {author} {\bibfnamefont {S.~B.}\ \bibnamefont {Cho}},\
  and\ \bibinfo {author} {\bibfnamefont {R.}~\bibnamefont {Mishra}},\
  }\bibfield  {title} {\bibinfo {title} {Multiferroism in iron-based
  oxyfluoride perovskites},\ }\href@noop {} {\bibfield  {journal} {\bibinfo
  {journal} {Inorganic Chemistry}\ }\textbf {\bibinfo {volume} {57}},\ \bibinfo
  {pages} {10616} (\bibinfo {year} {2018})}\BibitemShut {NoStop}%
\bibitem [{\citenamefont {Charles}\ \emph {et~al.}(2018)\citenamefont
  {Charles}, \citenamefont {Saballos},\ and\ \citenamefont
  {Rondinelli}}]{charles2018structural}%
  \BibitemOpen
  \bibfield  {author} {\bibinfo {author} {\bibfnamefont {N.}~\bibnamefont
  {Charles}}, \bibinfo {author} {\bibfnamefont {R.~J.}\ \bibnamefont
  {Saballos}},\ and\ \bibinfo {author} {\bibfnamefont {J.~M.}\ \bibnamefont
  {Rondinelli}},\ }\bibfield  {title} {\bibinfo {title} {Structural diversity
  from anion order in heteroanionic materials},\ }\href@noop {} {\bibfield
  {journal} {\bibinfo  {journal} {Chemistry of Materials}\ }\textbf {\bibinfo
  {volume} {30}},\ \bibinfo {pages} {3528} (\bibinfo {year}
  {2018})}\BibitemShut {NoStop}%
\bibitem [{\citenamefont {Yang}\ \emph {et~al.}(2011)\citenamefont {Yang},
  \citenamefont {Or{\'o}-Sol{\'e}}, \citenamefont {Rodgers}, \citenamefont
  {Jorge}, \citenamefont {Fuertes},\ and\ \citenamefont
  {Attfield}}]{yang2011anion}%
  \BibitemOpen
  \bibfield  {author} {\bibinfo {author} {\bibfnamefont {M.}~\bibnamefont
  {Yang}}, \bibinfo {author} {\bibfnamefont {J.}~\bibnamefont
  {Or{\'o}-Sol{\'e}}}, \bibinfo {author} {\bibfnamefont {J.~A.}\ \bibnamefont
  {Rodgers}}, \bibinfo {author} {\bibfnamefont {A.~B.}\ \bibnamefont {Jorge}},
  \bibinfo {author} {\bibfnamefont {A.}~\bibnamefont {Fuertes}},\ and\ \bibinfo
  {author} {\bibfnamefont {J.~P.}\ \bibnamefont {Attfield}},\ }\bibfield
  {title} {\bibinfo {title} {Anion order in perovskite oxynitrides},\
  }\href@noop {} {\bibfield  {journal} {\bibinfo  {journal} {Nature Chemistry}\
  }\textbf {\bibinfo {volume} {3}},\ \bibinfo {pages} {47} (\bibinfo {year}
  {2011})}\BibitemShut {NoStop}%
\bibitem [{\citenamefont {Pilania}\ \emph {et~al.}(2020)\citenamefont
  {Pilania}, \citenamefont {Ghosh}, \citenamefont {Hartman}, \citenamefont
  {Mishra}, \citenamefont {Stanek},\ and\ \citenamefont
  {Uberuaga}}]{pilania2020anion}%
  \BibitemOpen
  \bibfield  {author} {\bibinfo {author} {\bibfnamefont {G.}~\bibnamefont
  {Pilania}}, \bibinfo {author} {\bibfnamefont {A.}~\bibnamefont {Ghosh}},
  \bibinfo {author} {\bibfnamefont {S.~T.}\ \bibnamefont {Hartman}}, \bibinfo
  {author} {\bibfnamefont {R.}~\bibnamefont {Mishra}}, \bibinfo {author}
  {\bibfnamefont {C.~R.}\ \bibnamefont {Stanek}},\ and\ \bibinfo {author}
  {\bibfnamefont {B.~P.}\ \bibnamefont {Uberuaga}},\ }\bibfield  {title}
  {\bibinfo {title} {{Anion order in oxysulfide perovskites: origins and
  implications}},\ }\href@noop {} {\bibfield  {journal} {\bibinfo  {journal}
  {NPJ Computational Materials}\ }\textbf {\bibinfo {volume} {6}},\ \bibinfo
  {pages} {71} (\bibinfo {year} {2020})}\BibitemShut {NoStop}%
\bibitem [{\citenamefont {Krylov}\ \emph {et~al.}(2014)\citenamefont {Krylov},
  \citenamefont {Sofronova}, \citenamefont {Kolesnikova}, \citenamefont
  {Ivanov}, \citenamefont {Sukhovsky}, \citenamefont {Goryainov}, \citenamefont
  {Ivanenko}, \citenamefont {Shestakov}, \citenamefont {Kocharova},\ and\
  \citenamefont {Vtyurin}}]{krylov2014experimental}%
  \BibitemOpen
  \bibfield  {author} {\bibinfo {author} {\bibfnamefont {A.}~\bibnamefont
  {Krylov}}, \bibinfo {author} {\bibfnamefont {S.}~\bibnamefont {Sofronova}},
  \bibinfo {author} {\bibfnamefont {E.}~\bibnamefont {Kolesnikova}}, \bibinfo
  {author} {\bibfnamefont {Y.~N.}\ \bibnamefont {Ivanov}}, \bibinfo {author}
  {\bibfnamefont {A.}~\bibnamefont {Sukhovsky}}, \bibinfo {author}
  {\bibfnamefont {S.}~\bibnamefont {Goryainov}}, \bibinfo {author}
  {\bibfnamefont {A.}~\bibnamefont {Ivanenko}}, \bibinfo {author}
  {\bibfnamefont {N.}~\bibnamefont {Shestakov}}, \bibinfo {author}
  {\bibfnamefont {A.}~\bibnamefont {Kocharova}},\ and\ \bibinfo {author}
  {\bibfnamefont {A.}~\bibnamefont {Vtyurin}},\ }\bibfield  {title} {\bibinfo
  {title} {{Experimental and theoretical methods to study structural phase
  transition mechanisms in K$_3$WO$_3$F$_3$ oxyfluoride}},\ }\href@noop {}
  {\bibfield  {journal} {\bibinfo  {journal} {Journal of Solid State
  Chemistry}\ }\textbf {\bibinfo {volume} {218}},\ \bibinfo {pages} {32}
  (\bibinfo {year} {2014})}\BibitemShut {NoStop}%
\bibitem [{\citenamefont {Atuchin}\ \emph {et~al.}(2012)\citenamefont
  {Atuchin}, \citenamefont {Isaenko}, \citenamefont {Kesler}, \citenamefont
  {Lin}, \citenamefont {Molokeev}, \citenamefont {Yelisseyev},\ and\
  \citenamefont {Zhurkov}}]{atuchin2012exploration}%
  \BibitemOpen
  \bibfield  {author} {\bibinfo {author} {\bibfnamefont {V.}~\bibnamefont
  {Atuchin}}, \bibinfo {author} {\bibfnamefont {L.}~\bibnamefont {Isaenko}},
  \bibinfo {author} {\bibfnamefont {V.}~\bibnamefont {Kesler}}, \bibinfo
  {author} {\bibfnamefont {Z.}~\bibnamefont {Lin}}, \bibinfo {author}
  {\bibfnamefont {M.}~\bibnamefont {Molokeev}}, \bibinfo {author}
  {\bibfnamefont {A.}~\bibnamefont {Yelisseyev}},\ and\ \bibinfo {author}
  {\bibfnamefont {S.}~\bibnamefont {Zhurkov}},\ }\bibfield  {title} {\bibinfo
  {title} {{Exploration on anion ordering, optical properties and electronic
  structure in K$_3$WO$_3$F$_3$ elpasolite}},\ }\href@noop {} {\bibfield
  {journal} {\bibinfo  {journal} {Journal of Solid State Chemistry}\ }\textbf
  {\bibinfo {volume} {187}},\ \bibinfo {pages} {159} (\bibinfo {year}
  {2012})}\BibitemShut {NoStop}%
\bibitem [{\citenamefont {Tsujimoto}\ \emph {et~al.}(2018)\citenamefont
  {Tsujimoto}, \citenamefont {Juillerat}, \citenamefont {Zhang}, \citenamefont
  {Fujii}, \citenamefont {Yashima}, \citenamefont {Halasyamani},\ and\
  \citenamefont {zur Loye}}]{tsujimoto2018function}%
  \BibitemOpen
  \bibfield  {author} {\bibinfo {author} {\bibfnamefont {Y.}~\bibnamefont
  {Tsujimoto}}, \bibinfo {author} {\bibfnamefont {C.~A.}\ \bibnamefont
  {Juillerat}}, \bibinfo {author} {\bibfnamefont {W.}~\bibnamefont {Zhang}},
  \bibinfo {author} {\bibfnamefont {K.}~\bibnamefont {Fujii}}, \bibinfo
  {author} {\bibfnamefont {M.}~\bibnamefont {Yashima}}, \bibinfo {author}
  {\bibfnamefont {P.~S.}\ \bibnamefont {Halasyamani}},\ and\ \bibinfo {author}
  {\bibfnamefont {H.-C.}\ \bibnamefont {zur Loye}},\ }\bibfield  {title}
  {\bibinfo {title} {{Function of tetrahedral ZnS$_3$O building blocks in the
  formation of SrZn$_2$S$_2$O: a phase matchable polar oxysulfide with a large
  second harmonic generation response}},\ }\href@noop {} {\bibfield  {journal}
  {\bibinfo  {journal} {Chemistry of Materials}\ }\textbf {\bibinfo {volume}
  {30}},\ \bibinfo {pages} {6486} (\bibinfo {year} {2018})}\BibitemShut
  {NoStop}%
\bibitem [{\citenamefont {Oka}\ \emph {et~al.}(2014)\citenamefont {Oka},
  \citenamefont {Hirose}, \citenamefont {Kamisaka}, \citenamefont {Fukumura},
  \citenamefont {Sasa}, \citenamefont {Ishii}, \citenamefont {Matsuzaki},
  \citenamefont {Sato}, \citenamefont {Ikuhara},\ and\ \citenamefont
  {Hasegawa}}]{oka2014possible}%
  \BibitemOpen
  \bibfield  {author} {\bibinfo {author} {\bibfnamefont {D.}~\bibnamefont
  {Oka}}, \bibinfo {author} {\bibfnamefont {Y.}~\bibnamefont {Hirose}},
  \bibinfo {author} {\bibfnamefont {H.}~\bibnamefont {Kamisaka}}, \bibinfo
  {author} {\bibfnamefont {T.}~\bibnamefont {Fukumura}}, \bibinfo {author}
  {\bibfnamefont {K.}~\bibnamefont {Sasa}}, \bibinfo {author} {\bibfnamefont
  {S.}~\bibnamefont {Ishii}}, \bibinfo {author} {\bibfnamefont
  {H.}~\bibnamefont {Matsuzaki}}, \bibinfo {author} {\bibfnamefont
  {Y.}~\bibnamefont {Sato}}, \bibinfo {author} {\bibfnamefont {Y.}~\bibnamefont
  {Ikuhara}},\ and\ \bibinfo {author} {\bibfnamefont {T.}~\bibnamefont
  {Hasegawa}},\ }\bibfield  {title} {\bibinfo {title} {{Possible
  ferroelectricity in perovskite oxynitride SrTaO$_2$N epitaxial thin films}},\
  }\href@noop {} {\bibfield  {journal} {\bibinfo  {journal} {Scientific
  Reports}\ }\textbf {\bibinfo {volume} {4}},\ \bibinfo {pages} {4987}
  (\bibinfo {year} {2014})}\BibitemShut {NoStop}%
\bibitem [{\citenamefont {Zhang}\ \emph {et~al.}(2011)\citenamefont {Zhang},
  \citenamefont {Motohashi}, \citenamefont {Masubuchi},\ and\ \citenamefont
  {Kikkawa}}]{zhang2011local}%
  \BibitemOpen
  \bibfield  {author} {\bibinfo {author} {\bibfnamefont {Y.-R.}\ \bibnamefont
  {Zhang}}, \bibinfo {author} {\bibfnamefont {T.}~\bibnamefont {Motohashi}},
  \bibinfo {author} {\bibfnamefont {Y.}~\bibnamefont {Masubuchi}},\ and\
  \bibinfo {author} {\bibfnamefont {S.}~\bibnamefont {Kikkawa}},\ }\bibfield
  {title} {\bibinfo {title} {{Local anionic ordering and anisotropic
  displacement in dielectric perovskite SrTaO$_2$N}},\ }\href@noop {}
  {\bibfield  {journal} {\bibinfo  {journal} {Journal of the Ceramic Society of
  Japan}\ }\textbf {\bibinfo {volume} {119}},\ \bibinfo {pages} {581} (\bibinfo
  {year} {2011})}\BibitemShut {NoStop}%
\bibitem [{\citenamefont {Coles}\ \emph {et~al.}(2023)\citenamefont {Coles},
  \citenamefont {Falkowski}, \citenamefont {Geddes}, \citenamefont {P{\'e}rez},
  \citenamefont {Booth}, \citenamefont {Squires}, \citenamefont {O'Rourke},
  \citenamefont {McColl}, \citenamefont {Goodwin}, \citenamefont {Cussen} \emph
  {et~al.}}]{coles2023anion}%
  \BibitemOpen
  \bibfield  {author} {\bibinfo {author} {\bibfnamefont {S.~W.}\ \bibnamefont
  {Coles}}, \bibinfo {author} {\bibfnamefont {V.}~\bibnamefont {Falkowski}},
  \bibinfo {author} {\bibfnamefont {H.~S.}\ \bibnamefont {Geddes}}, \bibinfo
  {author} {\bibfnamefont {G.~E.}\ \bibnamefont {P{\'e}rez}}, \bibinfo {author}
  {\bibfnamefont {S.~G.}\ \bibnamefont {Booth}}, \bibinfo {author}
  {\bibfnamefont {A.~G.}\ \bibnamefont {Squires}}, \bibinfo {author}
  {\bibfnamefont {C.}~\bibnamefont {O'Rourke}}, \bibinfo {author}
  {\bibfnamefont {K.}~\bibnamefont {McColl}}, \bibinfo {author} {\bibfnamefont
  {A.~L.}\ \bibnamefont {Goodwin}}, \bibinfo {author} {\bibfnamefont {S.~A.}\
  \bibnamefont {Cussen}}, \emph {et~al.},\ }\bibfield  {title} {\bibinfo
  {title} {{Anion-polarisation-directed short-range-order in antiperovskite
  Li$_2$FeSO}},\ }\href@noop {} {\bibfield  {journal} {\bibinfo  {journal}
  {Journal of Materials Chemistry A}\ }\textbf {\bibinfo {volume} {11}},\
  \bibinfo {pages} {13016} (\bibinfo {year} {2023})}\BibitemShut {NoStop}%
\bibitem [{\citenamefont {Tsujimoto}\ \emph {et~al.}(2022)\citenamefont
  {Tsujimoto}, \citenamefont {Sugiyama}, \citenamefont {Ochi}, \citenamefont
  {Kuroki}, \citenamefont {Manuel}, \citenamefont {Khalyavin}, \citenamefont
  {Umegaki}, \citenamefont {M{\aa}nsson}, \citenamefont {Andreica},
  \citenamefont {Hara} \emph {et~al.}}]{tsujimoto2022impact}%
  \BibitemOpen
  \bibfield  {author} {\bibinfo {author} {\bibfnamefont {Y.}~\bibnamefont
  {Tsujimoto}}, \bibinfo {author} {\bibfnamefont {J.}~\bibnamefont {Sugiyama}},
  \bibinfo {author} {\bibfnamefont {M.}~\bibnamefont {Ochi}}, \bibinfo {author}
  {\bibfnamefont {K.}~\bibnamefont {Kuroki}}, \bibinfo {author} {\bibfnamefont
  {P.}~\bibnamefont {Manuel}}, \bibinfo {author} {\bibfnamefont {D.~D.}\
  \bibnamefont {Khalyavin}}, \bibinfo {author} {\bibfnamefont {I.}~\bibnamefont
  {Umegaki}}, \bibinfo {author} {\bibfnamefont {M.}~\bibnamefont
  {M{\aa}nsson}}, \bibinfo {author} {\bibfnamefont {D.}~\bibnamefont
  {Andreica}}, \bibinfo {author} {\bibfnamefont {S.}~\bibnamefont {Hara}},
  \emph {et~al.},\ }\bibfield  {title} {\bibinfo {title} {{Impact of mixed
  anion ordered state on the magnetic ground states of $S= 1/2$ square-lattice
  quantum spin antiferromagnets, Sr$_2$NiO$_3$Cl and Sr$_2$NiO$_3$F}},\
  }\href@noop {} {\bibfield  {journal} {\bibinfo  {journal} {Physical Review
  Materials}\ }\textbf {\bibinfo {volume} {6}},\ \bibinfo {pages} {114404}
  (\bibinfo {year} {2022})}\BibitemShut {NoStop}%
\bibitem [{\citenamefont {Brink}\ \emph {et~al.}(2000)\citenamefont {Brink},
  \citenamefont {Withers},\ and\ \citenamefont {Thompson}}]{brink2000electron}%
  \BibitemOpen
  \bibfield  {author} {\bibinfo {author} {\bibfnamefont {F.~J.}\ \bibnamefont
  {Brink}}, \bibinfo {author} {\bibfnamefont {R.~L.}\ \bibnamefont {Withers}},\
  and\ \bibinfo {author} {\bibfnamefont {J.~G.}\ \bibnamefont {Thompson}},\
  }\bibfield  {title} {\bibinfo {title} {{An electron diffraction and crystal
  chemical investigation of oxygen/fluorine ordering in rutile-type iron
  oxyfluoride, FeOF}},\ }\href@noop {} {\bibfield  {journal} {\bibinfo
  {journal} {Journal of Solid State Chemistry}\ }\textbf {\bibinfo {volume}
  {155}},\ \bibinfo {pages} {359} (\bibinfo {year} {2000})}\BibitemShut
  {NoStop}%
\bibitem [{\citenamefont {Chamberland}\ and\ \citenamefont
  {Sleight}(1967)}]{chamberland1967preparation}%
  \BibitemOpen
  \bibfield  {author} {\bibinfo {author} {\bibfnamefont {B.}~\bibnamefont
  {Chamberland}}\ and\ \bibinfo {author} {\bibfnamefont {A.}~\bibnamefont
  {Sleight}},\ }\bibfield  {title} {\bibinfo {title} {{Preparation of first-row
  transition metal oxyfluorides of the composition MOF}},\ }\href@noop {}
  {\bibfield  {journal} {\bibinfo  {journal} {Solid State Communications}\
  }\textbf {\bibinfo {volume} {5}},\ \bibinfo {pages} {765} (\bibinfo {year}
  {1967})}\BibitemShut {NoStop}%
\bibitem [{\citenamefont {Cumby}\ \emph {et~al.}(2018)\citenamefont {Cumby},
  \citenamefont {Burchell},\ and\ \citenamefont {Attfield}}]{cumby2018high}%
  \BibitemOpen
  \bibfield  {author} {\bibinfo {author} {\bibfnamefont {J.}~\bibnamefont
  {Cumby}}, \bibinfo {author} {\bibfnamefont {M.}~\bibnamefont {Burchell}},\
  and\ \bibinfo {author} {\bibfnamefont {J.}~\bibnamefont {Attfield}},\
  }\bibfield  {title} {\bibinfo {title} {{High pressure synthesis, crystal
  growth and magnetic properties of TiOF}},\ }\href@noop {} {\bibfield
  {journal} {\bibinfo  {journal} {Solid State Sciences}\ }\textbf {\bibinfo
  {volume} {80}},\ \bibinfo {pages} {35} (\bibinfo {year} {2018})}\BibitemShut
  {NoStop}%
\bibitem [{\citenamefont {Hart}\ and\ \citenamefont
  {Forcade}(2008)}]{hart2008algorithm}%
  \BibitemOpen
  \bibfield  {author} {\bibinfo {author} {\bibfnamefont {G.~L.}\ \bibnamefont
  {Hart}}\ and\ \bibinfo {author} {\bibfnamefont {R.~W.}\ \bibnamefont
  {Forcade}},\ }\bibfield  {title} {\bibinfo {title} {Algorithm for generating
  derivative structures},\ }\href@noop {} {\bibfield  {journal} {\bibinfo
  {journal} {Physical Review B}\ }\textbf {\bibinfo {volume} {77}},\ \bibinfo
  {pages} {224115} (\bibinfo {year} {2008})}\BibitemShut {NoStop}%
\bibitem [{\citenamefont {{\AA}ngqvist}\ \emph {et~al.}(2019)\citenamefont
  {{\AA}ngqvist}, \citenamefont {Mu{\~n}oz}, \citenamefont {Rahm},
  \citenamefont {Fransson}, \citenamefont {Durniak}, \citenamefont {Rozyczko},
  \citenamefont {Rod},\ and\ \citenamefont {Erhart}}]{aangqvist2019icet}%
  \BibitemOpen
  \bibfield  {author} {\bibinfo {author} {\bibfnamefont {M.}~\bibnamefont
  {{\AA}ngqvist}}, \bibinfo {author} {\bibfnamefont {W.~A.}\ \bibnamefont
  {Mu{\~n}oz}}, \bibinfo {author} {\bibfnamefont {J.~M.}\ \bibnamefont {Rahm}},
  \bibinfo {author} {\bibfnamefont {E.}~\bibnamefont {Fransson}}, \bibinfo
  {author} {\bibfnamefont {C.}~\bibnamefont {Durniak}}, \bibinfo {author}
  {\bibfnamefont {P.}~\bibnamefont {Rozyczko}}, \bibinfo {author}
  {\bibfnamefont {T.~H.}\ \bibnamefont {Rod}},\ and\ \bibinfo {author}
  {\bibfnamefont {P.}~\bibnamefont {Erhart}},\ }\bibfield  {title} {\bibinfo
  {title} {Icet--a python library for constructing and sampling alloy cluster
  expansions},\ }\href@noop {} {\bibfield  {journal} {\bibinfo  {journal}
  {Advanced Theory and Simulations}\ }\textbf {\bibinfo {volume} {2}},\
  \bibinfo {pages} {1900015} (\bibinfo {year} {2019})}\BibitemShut {NoStop}%
\bibitem [{\citenamefont {Kresse}\ and\ \citenamefont
  {Furthm{\"u}ller}(1996)}]{kresse1996efficient}%
  \BibitemOpen
  \bibfield  {author} {\bibinfo {author} {\bibfnamefont {G.}~\bibnamefont
  {Kresse}}\ and\ \bibinfo {author} {\bibfnamefont {J.}~\bibnamefont
  {Furthm{\"u}ller}},\ }\bibfield  {title} {\bibinfo {title} {Efficient
  iterative schemes for ab initio total-energy calculations using a plane-wave
  basis set},\ }\href@noop {} {\bibfield  {journal} {\bibinfo  {journal}
  {Physical Review B}\ }\textbf {\bibinfo {volume} {54}},\ \bibinfo {pages}
  {11169} (\bibinfo {year} {1996})}\BibitemShut {NoStop}%
\bibitem [{\citenamefont {Kresse}\ and\ \citenamefont
  {Joubert}(1999)}]{kresse1999ultrasoft}%
  \BibitemOpen
  \bibfield  {author} {\bibinfo {author} {\bibfnamefont {G.}~\bibnamefont
  {Kresse}}\ and\ \bibinfo {author} {\bibfnamefont {D.}~\bibnamefont
  {Joubert}},\ }\bibfield  {title} {\bibinfo {title} {From ultrasoft
  pseudopotentials to the projector augmented-wave method},\ }\href@noop {}
  {\bibfield  {journal} {\bibinfo  {journal} {Physical review B}\ }\textbf
  {\bibinfo {volume} {59}},\ \bibinfo {pages} {1758} (\bibinfo {year}
  {1999})}\BibitemShut {NoStop}%
\bibitem [{\citenamefont {Perdew}\ \emph {et~al.}(1996)\citenamefont {Perdew},
  \citenamefont {Burke},\ and\ \citenamefont
  {Ernzerhof}}]{perdew1996generalized}%
  \BibitemOpen
  \bibfield  {author} {\bibinfo {author} {\bibfnamefont {J.~P.}\ \bibnamefont
  {Perdew}}, \bibinfo {author} {\bibfnamefont {K.}~\bibnamefont {Burke}},\ and\
  \bibinfo {author} {\bibfnamefont {M.}~\bibnamefont {Ernzerhof}},\ }\bibfield
  {title} {\bibinfo {title} {Generalized gradient approximation made simple},\
  }\href@noop {} {\bibfield  {journal} {\bibinfo  {journal} {Physical Review
  Letters}\ }\textbf {\bibinfo {volume} {77}},\ \bibinfo {pages} {3865}
  (\bibinfo {year} {1996})}\BibitemShut {NoStop}%
\bibitem [{\citenamefont {Larsen}\ \emph {et~al.}(2017)\citenamefont {Larsen},
  \citenamefont {Mortensen}, \citenamefont {Blomqvist}, \citenamefont
  {Castelli}, \citenamefont {Christensen}, \citenamefont {Du{\l}ak},
  \citenamefont {Friis}, \citenamefont {Groves}, \citenamefont {Hammer},
  \citenamefont {Hargus} \emph {et~al.}}]{larsen2017atomic}%
  \BibitemOpen
  \bibfield  {author} {\bibinfo {author} {\bibfnamefont {A.~H.}\ \bibnamefont
  {Larsen}}, \bibinfo {author} {\bibfnamefont {J.~J.}\ \bibnamefont
  {Mortensen}}, \bibinfo {author} {\bibfnamefont {J.}~\bibnamefont
  {Blomqvist}}, \bibinfo {author} {\bibfnamefont {I.~E.}\ \bibnamefont
  {Castelli}}, \bibinfo {author} {\bibfnamefont {R.}~\bibnamefont
  {Christensen}}, \bibinfo {author} {\bibfnamefont {M.}~\bibnamefont
  {Du{\l}ak}}, \bibinfo {author} {\bibfnamefont {J.}~\bibnamefont {Friis}},
  \bibinfo {author} {\bibfnamefont {M.~N.}\ \bibnamefont {Groves}}, \bibinfo
  {author} {\bibfnamefont {B.}~\bibnamefont {Hammer}}, \bibinfo {author}
  {\bibfnamefont {C.}~\bibnamefont {Hargus}}, \emph {et~al.},\ }\bibfield
  {title} {\bibinfo {title} {The atomic simulation environment—a python
  library for working with atoms},\ }\href@noop {} {\bibfield  {journal}
  {\bibinfo  {journal} {Journal of Physics: Condensed Matter}\ }\textbf
  {\bibinfo {volume} {29}},\ \bibinfo {pages} {273002} (\bibinfo {year}
  {2017})}\BibitemShut {NoStop}%
\bibitem [{\citenamefont {Eyert}(2002{\natexlab{a}})}]{eyert2002metal}%
  \BibitemOpen
  \bibfield  {author} {\bibinfo {author} {\bibfnamefont {V.}~\bibnamefont
  {Eyert}},\ }\bibfield  {title} {\bibinfo {title} {{The metal-insulator
  transitions of VO$_2$: A band theoretical approach}},\ }\href@noop {}
  {\bibfield  {journal} {\bibinfo  {journal} {Annalen der Physik}\ }\textbf
  {\bibinfo {volume} {514}},\ \bibinfo {pages} {650} (\bibinfo {year}
  {2002}{\natexlab{a}})}\BibitemShut {NoStop}%
\bibitem [{\citenamefont {Eyert}(2002{\natexlab{b}})}]{eyert2002metal2}%
  \BibitemOpen
  \bibfield  {author} {\bibinfo {author} {\bibfnamefont {V.}~\bibnamefont
  {Eyert}},\ }\bibfield  {title} {\bibinfo {title} {{The metal-insulator
  transition of NbO$_2$: an embedded Peierls instability}},\ }\href@noop {}
  {\bibfield  {journal} {\bibinfo  {journal} {Europhysics Letters}\ }\textbf
  {\bibinfo {volume} {58}},\ \bibinfo {pages} {851} (\bibinfo {year}
  {2002}{\natexlab{b}})}\BibitemShut {NoStop}%
\bibitem [{Sup()}]{Supp}%
  \BibitemOpen
  \href@noop {} {}\bibinfo {note} {See Supplemental Material at [URL will be
  inserted by publisher] for additional computational details, phonon
  calculations, ELF for monoclinic VO$_2$.}\BibitemShut {Stop}%
\bibitem [{\citenamefont {Miller}\ and\ \citenamefont
  {Rondinelli}(2022)}]{Miller_2022}%
  \BibitemOpen
  \bibfield  {author} {\bibinfo {author} {\bibfnamefont {K.~D.}\ \bibnamefont
  {Miller}}\ and\ \bibinfo {author} {\bibfnamefont {J.~M.}\ \bibnamefont
  {Rondinelli}},\ }\bibfield  {title} {\bibinfo {title} {{Carrier-induced
  metal-insulator transition in trirutile MgTa$_2$O$_6$}},\ }\href
  {https://doi.org/10.1103/physrevmaterials.6.075007} {\bibfield  {journal}
  {\bibinfo  {journal} {Physical Review Materials}\ }\textbf {\bibinfo {volume}
  {6}},\ \bibinfo {pages} {075007} (\bibinfo {year} {2022})}\BibitemShut
  {NoStop}%
\bibitem [{\citenamefont {Schueller}\ \emph {et~al.}(2021)\citenamefont
  {Schueller}, \citenamefont {Oey}, \citenamefont {Miller}, \citenamefont
  {Wyckoff}, \citenamefont {Zhang}, \citenamefont {Zhang}, \citenamefont
  {Wilson}, \citenamefont {Rondinelli},\ and\ \citenamefont
  {Seshadri}}]{Schueller_2021}%
  \BibitemOpen
  \bibfield  {author} {\bibinfo {author} {\bibfnamefont {E.~C.}\ \bibnamefont
  {Schueller}}, \bibinfo {author} {\bibfnamefont {Y.~M.}\ \bibnamefont {Oey}},
  \bibinfo {author} {\bibfnamefont {K.~D.}\ \bibnamefont {Miller}}, \bibinfo
  {author} {\bibfnamefont {K.~E.}\ \bibnamefont {Wyckoff}}, \bibinfo {author}
  {\bibfnamefont {R.}~\bibnamefont {Zhang}}, \bibinfo {author} {\bibfnamefont
  {W.}~\bibnamefont {Zhang}}, \bibinfo {author} {\bibfnamefont {S.~D.}\
  \bibnamefont {Wilson}}, \bibinfo {author} {\bibfnamefont {J.~M.}\
  \bibnamefont {Rondinelli}},\ and\ \bibinfo {author} {\bibfnamefont
  {R.}~\bibnamefont {Seshadri}},\ }\bibfield  {title} {\bibinfo {title}
  {{AB$_2$X$_6$ Compounds and the Stabilization of Trirutile Oxides}},\ }\href
  {https://doi.org/10.1021/acs.inorgchem.1c01366} {\bibfield  {journal}
  {\bibinfo  {journal} {Inorganic Chemistry}\ }\textbf {\bibinfo {volume}
  {60}},\ \bibinfo {pages} {9224} (\bibinfo {year} {2021})}\BibitemShut
  {NoStop}%
\bibitem [{\citenamefont {Silvi}\ and\ \citenamefont
  {Savin}(1994)}]{silvi1994classification}%
  \BibitemOpen
  \bibfield  {author} {\bibinfo {author} {\bibfnamefont {B.}~\bibnamefont
  {Silvi}}\ and\ \bibinfo {author} {\bibfnamefont {A.}~\bibnamefont {Savin}},\
  }\bibfield  {title} {\bibinfo {title} {Classification of chemical bonds based
  on topological analysis of electron localization functions},\ }\href@noop {}
  {\bibfield  {journal} {\bibinfo  {journal} {Nature}\ }\textbf {\bibinfo
  {volume} {371}},\ \bibinfo {pages} {683} (\bibinfo {year}
  {1994})}\BibitemShut {NoStop}%
\bibitem [{\citenamefont {Koumpouras}\ and\ \citenamefont
  {Larsson}(2020)}]{Koumpouras2020}%
  \BibitemOpen
  \bibfield  {author} {\bibinfo {author} {\bibfnamefont {K.}~\bibnamefont
  {Koumpouras}}\ and\ \bibinfo {author} {\bibfnamefont {J.~A.}\ \bibnamefont
  {Larsson}},\ }\bibfield  {title} {\bibinfo {title} {Distinguishing between
  chemical bonding and physical binding using electron localization function
  (elf)},\ }\href {https://doi.org/10.1088/1361-648x/ab7fd8} {\bibfield
  {journal} {\bibinfo  {journal} {Journal of Physics: Condensed Matter}\
  }\textbf {\bibinfo {volume} {32}},\ \bibinfo {pages} {315502} (\bibinfo
  {year} {2020})}\BibitemShut {NoStop}%
\bibitem [{\citenamefont {Dudarev}\ \emph {et~al.}(1998)\citenamefont
  {Dudarev}, \citenamefont {Botton}, \citenamefont {Savrasov}, \citenamefont
  {Humphreys},\ and\ \citenamefont {Sutton}}]{dudarev1998electron}%
  \BibitemOpen
  \bibfield  {author} {\bibinfo {author} {\bibfnamefont {S.~L.}\ \bibnamefont
  {Dudarev}}, \bibinfo {author} {\bibfnamefont {G.~A.}\ \bibnamefont {Botton}},
  \bibinfo {author} {\bibfnamefont {S.~Y.}\ \bibnamefont {Savrasov}}, \bibinfo
  {author} {\bibfnamefont {C.}~\bibnamefont {Humphreys}},\ and\ \bibinfo
  {author} {\bibfnamefont {A.~P.}\ \bibnamefont {Sutton}},\ }\bibfield  {title}
  {\bibinfo {title} {{Electron-energy-loss spectra and the structural stability
  of nickel oxide: An LSDA+U study}},\ }\href@noop {} {\bibfield  {journal}
  {\bibinfo  {journal} {Physical Review B}\ }\textbf {\bibinfo {volume} {57}},\
  \bibinfo {pages} {1505} (\bibinfo {year} {1998})}\BibitemShut {NoStop}%
\bibitem [{\citenamefont {Trimarchi}\ \emph {et~al.}(2018)\citenamefont
  {Trimarchi}, \citenamefont {Wang},\ and\ \citenamefont
  {Zunger}}]{PhysRevB.97.035107}%
  \BibitemOpen
  \bibfield  {author} {\bibinfo {author} {\bibfnamefont {G.}~\bibnamefont
  {Trimarchi}}, \bibinfo {author} {\bibfnamefont {Z.}~\bibnamefont {Wang}},\
  and\ \bibinfo {author} {\bibfnamefont {A.}~\bibnamefont {Zunger}},\
  }\bibfield  {title} {\bibinfo {title} {{Polymorphous band structure model of
  gapping in the antiferromagnetic and paramagnetic phases of the Mott
  insulators MnO, FeO, CoO, and NiO}},\ }\href
  {https://doi.org/10.1103/PhysRevB.97.035107} {\bibfield  {journal} {\bibinfo
  {journal} {Phys. Rev. B}\ }\textbf {\bibinfo {volume} {97}},\ \bibinfo
  {pages} {035107} (\bibinfo {year} {2018})}\BibitemShut {NoStop}%
\bibitem [{\citenamefont {Harada}\ \emph {et~al.}(2019)\citenamefont {Harada},
  \citenamefont {Charles}, \citenamefont {Poeppelmeier},\ and\ \citenamefont
  {Rondinelli}}]{harada2019heteroanionic}%
  \BibitemOpen
  \bibfield  {author} {\bibinfo {author} {\bibfnamefont {J.~K.}\ \bibnamefont
  {Harada}}, \bibinfo {author} {\bibfnamefont {N.}~\bibnamefont {Charles}},
  \bibinfo {author} {\bibfnamefont {K.~R.}\ \bibnamefont {Poeppelmeier}},\ and\
  \bibinfo {author} {\bibfnamefont {J.~M.}\ \bibnamefont {Rondinelli}},\
  }\bibfield  {title} {\bibinfo {title} {Heteroanionic materials by design:
  progress toward targeted properties},\ }\href@noop {} {\bibfield  {journal}
  {\bibinfo  {journal} {Advanced Materials}\ }\textbf {\bibinfo {volume}
  {31}},\ \bibinfo {pages} {1805295} (\bibinfo {year} {2019})}\BibitemShut
  {NoStop}%
\bibitem [{\citenamefont {Hiroi}(2015)}]{hiroi2015structural}%
  \BibitemOpen
  \bibfield  {author} {\bibinfo {author} {\bibfnamefont {Z.}~\bibnamefont
  {Hiroi}},\ }\bibfield  {title} {\bibinfo {title} {{Structural instability of
  the rutile compounds and its relevance to the metal--insulator transition of
  VO$_2$}},\ }\href@noop {} {\bibfield  {journal} {\bibinfo  {journal}
  {Progress in Solid State Chemistry}\ }\textbf {\bibinfo {volume} {43}},\
  \bibinfo {pages} {47} (\bibinfo {year} {2015})}\BibitemShut {NoStop}%
\end{thebibliography}%
\end{document}